\documentclass{llncs}
\usepackage{amsfonts}
\usepackage{latexsym}
\usepackage{amssymb}
\usepackage[dvips]{graphicx}
\usepackage{color}

\newtheorem{Propriete}{\bfseries Property}

\newcounter{CptrInvariant}
\newcommand{\CorpsInvariant}[1]{%
  \vskip 5pt%
    \refstepcounter{CptrInvariant}%
    \noindent $\!\!\!$\vrule width 0.5pt~%
    \begin{minipage}{\textwidth}%
      \noindent \textbf{Invariant~(#1)~}%
      \vspace*{-8pt}%
      \begin{center}%
        \itshape%
        \begin{tabular}{ll}%
}
\newenvironment{Invariant}[1]
               {\CorpsInvariant{#1}}
               {\end{tabular}
                \end{center}
                \end{minipage}
                \mdseries
                \vskip 5pt
               }

\newcounter{CptrProp}
\newcommand{\CorpsProp}[1]{
  \vskip 5pt
    \refstepcounter{CptrProp}
    \noindent $\!\!\!\!\!\!$ \vrule width 0.5pt ~ 
    \begin{minipage}{\textwidth}
      \noindent \textbf{#1:~}
      \vspace{-8pt}
      \begin{center}
        \itshape
        \begin{tabular}{ll}
}
\newenvironment{ConstantsProperties}[1]
               {\CorpsProp{#1} 
                }
               {\end{tabular}
                \end{center}
                \end{minipage}
                \mdseries
                \vskip 5pt
               }

\newcounter{CptrAssertion}
\newcommand{\CorpsAssertion}[1]{
  \vskip 5pt
    \refstepcounter{CptrAssertion}
    \noindent $\!\!\!\!\!\!$ \vrule width 0.5pt ~ 
    \begin{minipage}{\textwidth}
      \noindent \textbf{Assertion \theCptrAssertion ~{\bf (#1)}~}
      \vspace{-8pt}
      \begin{center}
        \itshape
        \begin{tabular}{ll}
}
\newenvironment{Assertion}[1]
               {\CorpsAssertion{#1} 
                }
               {\end{tabular}
                \end{center}
                \end{minipage}
                \mdseries
                \vskip 5pt
               }

\newcommand{\SP}{\mathit{SP}}
\newcommand{\Event}{\mathit{Net}}

\newcommand{\CONFLICT}{\mathit{CONFLICT}}
\newcommand{\FAIL}{\mathit{FAIL}}
\newcommand{\ConfigurationDomain}{\mathit{KnownNet}}

\newcommand{\USERS}{\mathit{USERS}}
\newcommand{\SERVICES}{\mathit{SERVICES}}
\newcommand{\DAEMONS}{\mathit{DAEMONS}}
\newcommand{\TerminalServers}{\mathit{TerminalServers}}
\newcommand{\HOSTS}{\mathit{HOSTS}}
\newcommand{\PORTS}{\mathit{PORTS}}
\newcommand{\INTERFACES}{\mathit{INTERFACES}}

\newcommand{\Users}{\mathit{Users}}
\newcommand{\Services}{\mathit{Services}}
\newcommand{\Daemons}{\mathit{Daemons}}
\newcommand{\Hosts}{\mathit{Hosts}}
\newcommand{\Interfaces}{\mathit{Interfaces}}
\newcommand{\Ports}{\mathit{Ports}}

\newcommand{\Monitored}{\mathit{Monitored}}
\newcommand{\Refines}{\mathit{Represents}}

\newcommand{\checkevent}{\mathit{Check\_event}}
\newcommand{\eventfilter}{\mathit{Event\_filter}}
\newcommand{\getstatus}{\mathit{Get\_status}}

\newcommand{\Conflict}{\mathit{Conflict}}
\newcommand{\Fail}{\mathit{Fail}}
\newcommand{\Pass}{\mathit{Pass}}

\newcommand{\EVar}{\mathit{Obs_{Event}}}
\newcommand{\ERes}{\mathit{Obs_{Status}}}

\newcommand{\provide}{\mathit{Provide}}
\newcommand{\usedby}{\mathit{Used\_By}}
\newcommand{\runon}{\mathit{Run\_on}}
\newcommand{\hosting}{\mathit{Hosting}}

\newcommand{\isoutofSP}{\mathit{Is\_e3\_Out\_Of\_\SP}}
\newcommand{\isinSP}{\mathit{Is\_e3\_In\_\SP}}
\newcommand{\WriteFail}{\mathit{WriteFail}}
\newcommand{\WriteConflict}{\mathit{WriteConflict}}

\newcommand{\isInConfDom}{\mathit{Is\_In\_\ConfigurationDomain}}

\newcommand{\Status}{\mathit{Status}}

\begin{document}


\newcommand{\WP}{${\cal WP}$}

\newcommand{\raf}{{\sqsubseteq}}	
\newcommand{\rafe}{{\leadsto}}	

\newcommand{\B}{{\sf B}}	
\newcommand{\Z}{{\sf Z}}	
\newcommand{\TLA}{{\sf TLA}}	
\newcommand{\ab}{{\it AtelierB}}	
\newcommand{\Bc}{\B\ classique}	
\newcommand{\Bl}{\B\ logiciel}	
\newcommand{\Be}{\B\ \'e\-v\'e\-ne\-men\-tiel}	
\newcommand{\Bes}{\B\ \'e\-v\'e\-ne\-men\-tiels}	
\newcommand{\Bs}{\B\ syst\`eme}	


\newcommand{\incl}{{\sc includes}}	
\newcommand{\uses}{{\sc uses}}	
\newcommand{\sees}{{\sc sees}}	
\newcommand{\impo}{{\sc imports}}	
\newcommand{\prom}{{\sc promotes}}	
\newcommand{\exte}{{\sc extends}}	
\newcommand{\refines}{{\sc refines}}	
\newcommand{\localop}{{\sc local$\_$operations}}	


\newcommand{\sys}{{\sc system}}	
\newcommand{\mach}{{\sc machine}}	
\newcommand{\ctrs}{{\sc constraints}}	
\newcommand{\defs}{{\sc definitions}}	
\newcommand{\sets}{{\sc sets}}	
\newcommand{\values}{{\sc values}}	
\newcommand{\cons}{{\sc constants}}	
\newcommand{\abstrcons}{{\sc abstract$\_$constants}}	
\newcommand{\concrcons}{{\sc concrete$\_$constants}}	
\newcommand{\prop}{{\sc properties}}	
\newcommand{\var}{{\sc variables}}	
\newcommand{\abstrvar}{{\sc abstract$\_$variables}}	
\newcommand{\concrvar}{{\sc concrete$\_$variables}}	
\newcommand{\inv}{{\sc invariant}}	
\newcommand{\variant}{{\sc variant}}	
\newcommand{\dyn}{{\sc dynamics}}	
\newcommand{\assert}{{\sc assertions}}	
\newcommand{\init}{{\sc initialisation}}
\newcommand{\oper}{{\sc operations}}	
\newcommand{\even}{{\sc events}}	
\newcommand{\refi}{{\sc refinement}}	
\newcommand{\impl}{{\sc implementation}}
\newcommand{\modal}{{\sc modalities}}	
\newcommand{\bend}{{\sc end}}		


\newcommand{\bbegin}{{\sc begin }}	
\newcommand{\bpre}{{\sc pre }}	

\newcommand{\bif}{{\sc if }}	
\newcommand{\bthen}{{\sc then}}	
\newcommand{\belse}{{\sc else }}	
\newcommand{\belsif}{{\sc elsif }}	
\newcommand{\bcase}{{\sc case }}	
\newcommand{\bof}{{\sc of }}	
\newcommand{\beither}{{\sc either }}	
\newcommand{\bor}{{\sc or }}	
\newcommand{\bselect}{{\sc select }}	
\newcommand{\bleadsto}{{\sc leadsto }}	
\newcommand{\bwhen}{{\sc when }}	
\newcommand{\bchoice}{{\sc choice }}	
\newcommand{\bwhile}{{\sc while }}	
\newcommand{\bdo}{{\sc do }}	
\newcommand{\binvariant}{{\sc invariant }}	
\newcommand{\bvariant}{{\sc variant }}	
\newcommand{\buntil}{{\sc until}}	

\newcommand{\bvar}{{\sc var }}		
\newcommand{\bany}{{\sc any }}		
\newcommand{\bwhere}{{\sc where }}	
\newcommand{\bin}{{\sc in }}		
\newcommand{\blet}{{\sc let }}		
\newcommand{\bbe}{{\sc be }}		

\newcommand{\bbool}{\mathsf{bool}}	
\newcommand{\bskip}{\mathsf{skip}}	


\newcommand{\logand}{~\wedge~}		
\newcommand{\logor}{~\vee~}		
\newcommand{\logimpli}{\Rightarrow}	
\newcommand{\logequiv}{\Leftrightarrow}	
\newcommand{\bdef}{~\hat =~}		


\newcommand{\bguard}{\Longrightarrow}	
\newcommand{\bch}{~[\!]~}		
\newcommand{\ba}{@}		
\newcommand{\bpt}{\cdot}		
\newcommand{\notfree}{\backslash}	
\newcommand{\bres}{\longleftarrow}	
\newcommand{\pv}{~\mathbf{;}~}		
\newcommand{\mapplet}{\mapsto}


\newcommand{\btrm}{\mathsf{trm}}	
\newcommand{\babt}{\mathsf{abt}}	
\newcommand{\bmir}{\mathsf{mir}}	
\newcommand{\bfis}{\mathsf{fis}}	
\newcommand{\bprd}{\mathsf{prd}}	
\newcommand{\bgar}{\mathsf{garde}}	

\newcommand{\rpre}{\mathsf{pre}}	
\newcommand{\rrel}{\mathsf{rel}}	
\newcommand{\rdom}{\mathsf{dom}}	
\newcommand{\rran}{\mathsf{ran}}	
\newcommand{\rstr}{\mathsf{str}}	
\newcommand{\rset}{\mathsf{set}}	
\newcommand{\rid}{\mathsf{id}}	
\newcommand{\rprj}{\mathsf{prj}}	


\newcommand{\domr}{\lhd}  	
\newcommand{\adomr}{\lhd\!\!\!\!\!\!-}  
\newcommand{\bpow}{{\mathbb P}} 	
\newcommand{\bpownv}{{\mathbb P}_1} 	
\newcommand{\bfin}{{\mathbb F}} 	
\newcommand{\bfinnv}{{\mathbb F}_1} 	
\newcommand{\brel}{\leftrightarrow}	
\newcommand{\btot}{\rightarrow}	
\newcommand{\bpar}{\rightarrow\!\!\!\!\!\shortmid~~}	
\newcommand{\bover}{~\mbox{{\large $<$}}\!\!{\mbox{+}~}}
\newcommand{\bssdom}{\lhd\!\!\!\!\!-}
\newcommand{\brestdom}{\lhd}

\newcommand{\bsscodom}{\rhd\!\!\!\!\!-}
\newcommand{\brestcodom}{\rhd}

\newcommand{\card}{\mathsf{card}}	
\newcommand{\bmin}{\mathsf{min}}	
\newcommand{\bmax}{\mathsf{max}}	
\newcommand{\injpar}{\rightarrowtail\!\!\!\!\!\shortmid~~}
\newcommand{\injective}{\rightarrowtail}
\newcommand{\surpar}{\rightarrow\!\!\!\!\!\rightarrow\!\!\!\!\!\shortmid~~}
\newcommand{\surjective}{\rightarrow\!\!\!\!\!\rightarrow}
\newcommand{\bijpar}{\rightarrowtail\!\!\!\!\!\rightarrow\!\!\!\!\!\shortmid~~}
\newcommand{\bijective}{\rightarrowtail\!\!\!\!\!\rightarrow}

\newcommand{\F}{{\cal F}}	
\newcommand{\N}{{\mathbb N}}	

\newcommand{\nat}{\mathsf{NAT}}	
\newcommand{\natpos}{\mathsf{NAT}_1}	
\newcommand{\intgen}{\mathsf{INT}}	
\newcommand{\intneg}{\mathsf{INT}_1}	
\newcommand{\caract}{\mathsf{CHAR}}	
\newcommand{\bool}{\mathsf{BOOL}}	
\newcommand{\true}{\mathsf{true}}	
\newcommand{\false}{\mathsf{false}}	
\newcommand{\btrue}{\mathsf{btrue}}	
\newcommand{\bfalse}{\mathsf{bfalse}}	
\newcommand{\bnon}{\mathsf{not}}	
\newcommand{\bou}{\mathsf{or}}	
\newcommand{\bet}{\mathsf{and}}	
\newcommand{\minint}{\mathsf{minint}}	
\newcommand{\maxint}{\mathsf{maxint}}	

\newcommand{\shaut}{\uparrow}	
\newcommand{\sbas}{\downarrow}	
\newcommand{\sapp}{\frown}	


\newcommand{\bigor}{\mathsf{OR}}	
\newcommand{\dirp}[1]{\otimes_{#1}}	

\newcommand{\biter}{{\sc iterate}}
\newcommand{\bunion}{{\sc union}}
\newcommand{\bUnion}{{\sc UNION}}
\newcommand{\bimpl}{\Rightarrow}




\def\machinebox#1{\centerline{\hbox{\fbox{\parbox[t]{4cm}{
	\vspace*{-3ex}
	\begin{tabbing}
	X\=X\=X\=X\=X\=X\=X\=X\=X\=X\=X\=X\=X\= \kill
	#1
	\end{tabbing}
	\vspace*{-3ex}
	}}}}
}


\def\machinesbox#1{\hbox{\fbox{\parbox[t]{4cm}{
        \vspace*{-3ex}
        \begin{tabbing}
        X\=X\=X\=X\=X\=XX\=XX\=XX\=XX\=XX\=XX\=XX\=XX\= \kill
        #1
        \end{tabbing}
        \vspace*{-3ex}
        }}}
}


\def\machinesboxSansBoite#1{\parbox[t]{4cm}{
        \vspace*{-3ex}
        \begin{tabbing}
        X\=X\=X\=X\=X\=XX\=XX\=XX\=XX\=XX\=XX\=XX\=XX\= \kill
        #1
        \end{tabbing}
        \vspace*{-3ex}
        }
}

\def\machineboxGrisee#1{\centerline{%
  \psframebox[fillstyle=solid,fillcolor=GrisClair,framesep=5pt]{%
    \parbox[t]{4cm}{
	\vspace*{-3ex}
	\begin{tabbing}
	X\=X\=X\=X\=X\=X\=X\=X\=X\=X\=X\=X\=X\= \kill
	#1
	\end{tabbing}
	\vspace*{-3ex}
	}}%
   }%
}

%
%
%



\newcommand{\clearsy}{ClearSy}
\newcommand{\java}{{\bf Java}}	
\newcommand{\javacard}{{\bf JavaCard}}
\newcommand{\bom}{{\bf \B OM}}
\newcommand{\bob}{{\bf Bo\B}}
\newcommand{\eden}{{\bf EDEN}}
\newcommand{\cc}{{\bf CC}}
\newcommand{\CC}{Critères Communs}
\newcommand{\toe}{{\bf TOE}}
\newcommand{\GS}{{\sf G\'en\'eSyst}}	
\newcommand{\jbt}{{\sf jBTools}}	
\newcommand{\csptob}{{\sf CSP2B}}
\newcommand{\bztt}{{\sf BZ-TT}}
\newcommand{\ProB}{{\sf ProB}}
\newcommand{\cadp}{{\sf CADP}}	
\newcommand{\graphviz}{{\sf GraphViz}}
\newcommand{\uml}{{\bf UML}}
\newcommand{\ocl}{{\bf OCL}}
\newcommand{\pltl}{{\bf PLTL}}
\newcommand{\unity}{{\bf UNITY}}

\newcommand{\demoney}{{\sf DEMONEY}}	

\newcommand{\ste}{sys\-t\`e\-me de tran\-si\-tions \'e\-ti\-que\-t\'ees}
\newcommand{\stes}{sys\-t\`e\-mes de tran\-si\-tions \'e\-ti\-que\-t\'ees}
\newcommand{\Ste}{Sys\-t\`e\-me de tran\-si\-tions \'e\-ti\-que\-t\'ees}
\newcommand{\Stes}{Sys\-t\`e\-mes de tran\-si\-tions \'e\-ti\-que\-t\'ees}
\newcommand{\STE}{\textbf{STE}}

\newcommand{\stc}{Statechart}
\newcommand{\stcs}{Statecharts}
\newcommand{\Stc}{Statechart}
\newcommand{\Stcs}{Statecharts}

\newcommand{\et}{diagramme d'\'etat/tran\-si\-tion}
\newcommand{\ets}{diagrammes d'\'etat/tran\-si\-tion}
\newcommand{\Et}{Diagramme d'\'etat/tran\-si\-tion}
\newcommand{\Ets}{Diagrammes d'\'etat/tran\-si\-tion}

\newcommand{\stepre}{\mathsf{Pre}}		
\newcommand{\stepret}{\widetilde \stepre}	
\newcommand{\stepost}{\mathsf{Post}}	
\newcommand{\stepostt}{\widetilde \stepost}	

\newcommand{\guard}{\mathsf{Guard}}
\newcommand{\garde}{\mathsf{Garde}}
\newcommand{\action}{\mathsf{Action}}
\newcommand{\sepGarde}{]~[}	
\newcommand{\traces}{\mathsf{Traces}}

\newcommand{\si}{système d'information}
\newcommand{\sis}{systèmes d'information}
\newcommand{\Si}{Système d'information}
\newcommand{\Sis}{Systèmes d'information}
\newcommand{\SI}{Système d'Information}
\newcommand{\SIs}{Systèmes d'Information}
\newcommand{\PP}{{\bf PP}}
\newcommand{\pp}{profil de protection}
\newcommand{\pps}{profils de protection}

\newcommand{\bytecode}{Bytecode}

\title{Security Policy Enforcement Through Refinement Process}

\author{Nicolas Stouls\thanks{Work supported by CNRS and ST-Microelectronics by the way of a doctoral grant.} \and Marie-Laure Potet}
\institute{Laboratoire Logiciels Syst\`emes R\'eseaux - LSR-IMAG - Grenoble, France\\
           \email{\{Nicolas.Stouls, Marie-Laure.Potet\}@imag.fr}}

\maketitle

\begin{abstract}

%
%

  In the area of networks, a common method to enforce a security policy expressed in a high-level language is based on an ad-hoc and manual rewriting process~\cite{WooLam}. We argue that it is possible to build a formal link between concrete and abstract terms, which can be dynamically computed from the environment data. In order to progressively introduce configuration data and then simplify the proof obligations, we use the \B\ refinement process. 
  We present a case study modeling a network monitor. 
  This program, described by refinement following the layers of the TCP/IP suite protocol, has to warn for all observed events which do not respect the security policy. To design this model, we use the event-\B\ method because it is suitable for modeling network concepts.

  This work has been done within the framework of the POTESTAT\footnote{
    Security policies: test directed analysis of open networks systems.\\http://www-lsr.imag.fr/POTESTAT/
  } project~\cite{potestat}, based on the research of network testing methods from a high-level security policy.

  \keywordname\ Security policy enforcement, refinement, TCP/IP layers.
\end{abstract}

  \section{Introduction}

  The separation between {\it policies} and {\it mechanisms} is considered as a main specification principle in security. The policy describes the authorized actions while the mechanism is the method to implement the policy~\cite{WooLam,Masullo93}. Those two concepts do not have the same abstraction level. The classical process to enforce a policy consists of manually rewriting the policy in the same terms as the mechanism, with ad-hoc methods. 
  We argue that a policy can be formally enforced in a mechanism by gradually building, through a refinement process, a link between abstract and concrete terms. 
  We propose to design a specification with the same abstraction level as the policy and to refine it to obtain the concrete mechanism. In the case of critical software, using an abstract specification is, for example, required for test and audit processes or for certification according to the Common Criteria~\cite{CC}. 

%
  
  To illustrate our approach, we describe a network security software which has to enforce an abstract security policy in a TCP/IP network. 
  Modern TCP/IP networks are heterogeneous and distributed, and their management becomes more and more complex. Thus, the use of an abstract security policy can give a global and comprehensive view of a network security~\cite{Sloman94}. We choose to focus more specifically on an access control policy because it is the main concept in network security \cite{Denning79,Sandhu92}.

  We aim at designing a monitor, which warns if an action, forbidden by the policy, is observed on the network. In order to achieve that, we use the event-\B\ method~\cite{Abrial-extending} for modeling network concepts. 

%



  The next section is an overview of the event-\B\ method.
  Section 3 introduces networks and their security policy concepts. 
  Then, Section~4 presents our approach, 
  Section~5 describes our method based on the refinement process. 
  Section~6 is a presentation of the case study.
  Finally, we conclude by comparing this work to related ones and by giving some prospects.


\vspace*{-5pt}
\section{Event-\B }
\vspace*{-5pt}

  The \B\ method \cite{BBook} is a formal development method as well as a specification language. \B\ components can be refined and implemented. The correctness of models and refinements can be validated by proof obligations.

  Event-\B\ \cite{Abrial-extending} is an extension of the \B\ language where models are described by events instead of operations. The most abstract component is called {\it system}. 
  Each event is composed by a guard $G$ and an action $T$ such that 
  if $G$ is enabled, then $T$ can be executed. If several guards are enabled at the same time then the triggered event is 
  chosen in a nondeterministic way.

  Through the refinement process, data representation can be changed. The gluing invariant describes the relationship between abstract and concrete variables. If an event $e_A$ is refined by an event $e_R$, then the refinement guard has to imply the abstract one. Moreover, some events can be introduced during the refinement process (refining the $\bskip$ event), according to the same principles as the {\it stuttering} in TLA~\cite{tla}. Due to the guard strengthening through refinement process, we have to prove that there is always at least one enabled event (no dead-lock) and that new events do not introduce live-locks.

  To conclude, Table~\ref{Tab-OperateursEnsemblistes} defines the set notations which are used thereafter and Table~\ref{Tab-SubstGeneralisees} summarizes generalised substitutions.

   \begin{table}[ht]
     \vskip -15pt
     \begin{center}
      \begin{tabular}{|lcl|}
        \hline
        Operator && Meaning\\
        \hline
        \hline
        $A \brel B$   &~$\bdef$~& $\{R ~|~ R \subseteq A \times B\}$\\
        $\rdom(R)$    &~$\bdef$~& $\{a ~|~ \exists b \cdot ((a,b) \in R)\}$\\
        $\rran(R)$    &~$\bdef$~& $\{b ~|~ \exists a \cdot ((a,b) \in R)\}$\\
        $R[A]$        &~$\bdef$~& $\{b ~|~ \exists a \cdot (a \in A \logand (a,b) \in R)\}$\\
        $R^{-1}$      &~$\bdef$~& $\{(b,a) ~|~ (a,b) \in R\}$\\
        $R_1~;~R_2$   &~$\bdef$~& $\{(a,c) ~|~ \exists b \cdot ((a,b) \in R_1 \logand (b,c) \in R_2)\}$\\
        $R_1~||~R_2$  &~$\bdef$~& $\{((a,b),(c,d)) ~|~ (a,c) \in R_1 \logand (b,d) \in R_2\}$ \\
         $R \brestcodom B$ &~$\bdef$~& $\{(a,b) ~|~ (a,b) \in R \logand b\in B\}$\\
        $A \bpar B$   &~$\bdef$~& $\{F ~|~ F \in A \brel B \logand \forall (b_1,b_2) \cdot ((a,b_1) \in F \logand (a,b_2) \in F \bimpl b_1=b_2)\}$\\
  
        \hline
      \end{tabular}
     \end{center}
     \vskip -5pt
     \caption{Used sets operators}
     \label{Tab-OperateursEnsemblistes}
     \vskip -10pt
   \end{table}
   
   \begin{table}[h]
     \small
     \begin{center}
       \begin{tabular}{|@{~~}l|@{~~}l|@{~~}l|}
         \hline
         Substitution & Syntactical notation & Mathematical notation\\
         \hline
         \hline
         Do nothing & $\bskip$ & $\bskip$\\
         Assignment & $x:=E$ & $x:=E$\\
         Unbounded choice ~~& \bany $z$ \bwhere P \bthen\ $T$ \bend~~  & $@z \cdot (P \bimpl T)$\\
         Condition & \bif $P$ \bthen\ $T_1$ \belse\ $T_2$ \bend & $P \bguard T_1 \bch \neg P \bguard T_2~~$\\
         \hline
       \end{tabular}
     \end{center}
     \vskip -5pt
     \caption{Used primitives substitutions}
     \label{Tab-SubstGeneralisees}
     \vskip -45pt
   \end{table}

\vspace*{-2pt}
\section{Introduction to Networks and their Security Policies}
    \vspace*{-2pt}
  \subsection{The TCP/IP Protocol Suite}
    \vspace*{-2pt}
  
  Computer networks use a standard connection model, called OSI (Open Systems Interconnection)~\cite{ModeleOSI}, composed of seven layers. The TCP/IP protocol suite implements this model but is described with only four layers: application, transport, network and link. Each of these layers plays a particular role:

  \vspace*{-5pt}
  \begin{itemize}
    \item The {\it Application} layer is the interface between the applications and the network (client-side protocol).
    \item The {\it Transport} layer manages the host-to-host communications, but not the route between them (peer-to-peer networks).
    \item The {\it Network} layer manages the route between networks by selecting the network interface to use and the first router.
    \item The {\it Link} layer performs the signal translation (analogic/numeric) and synchronizes the data transmission. This layer is most often provided by the hardware. 
    Therefore, it is not considered in the following sections.
  \end{itemize}


  \begin{figure}[ht]
    \vspace*{-25pt}
    \centering
    \includegraphics[width=11cm]{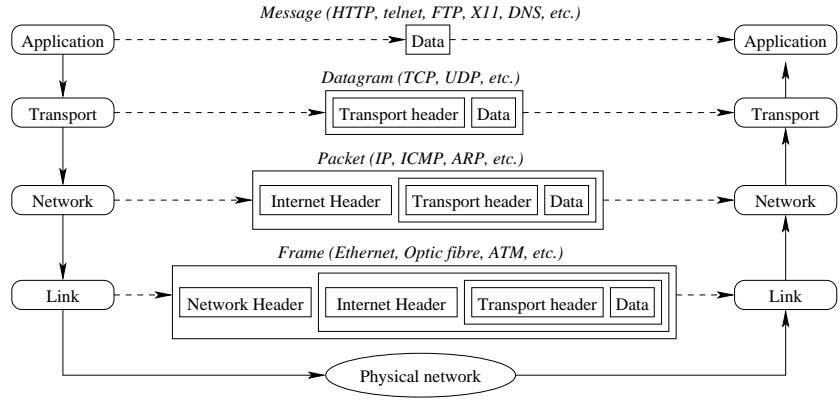}
    \vspace*{-10pt}
    \caption{Layers of the TCP/IP suite with some examples of protocols.}
    \label{ModeleTCPIP}
    \vspace*{-15pt}
  \end{figure}

  Communications using TCP/IP protocol are composed of protocols for each layer in the suite. For example, a TCP datagram (Transport layer) is contained in the data field of an IP packet (Network layer). Figure~\ref{ModeleTCPIP} shows an example of communication using TCP/IP.

    \vspace*{-10pt}

  \subsection{Network Security Policies}

\begin{figure}[ht]
    \vskip -25pt
    \centerline{
      \begin{tabular}{cccc}%
        ~~\includegraphics[width=3cm]{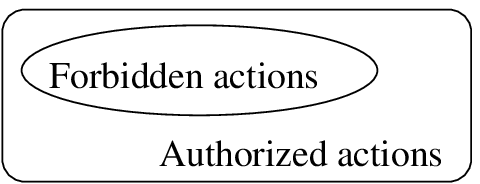}~~
        &~~\includegraphics[width=3cm]{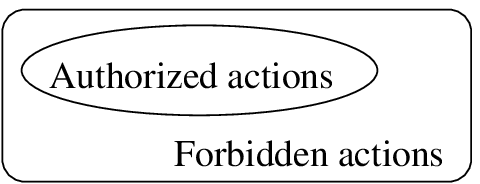}~~
        &~~\includegraphics[width=3.5cm]{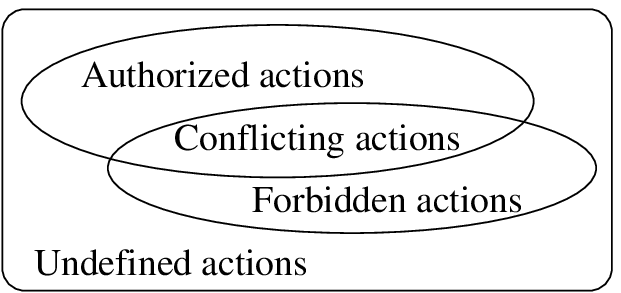}~~\\
        A. Open policy
        &B. Closed policy
        &C. Both policies
      \end{tabular}
    }
    \vskip -7pt
  
    \caption{Example of open, closed and both policies.}
    \label{ExempleDesDifferentesPolitiques}
    \vskip -15pt
  \end{figure}

  In the area of networks, security is mainly expressed in terms of access rights. An access control policy is defined on a set of actions by a set of rules. These rules determine, for each action, whether the action is authorized or not.
  Among the various types of access control policies~\cite{Jajodia97,Lunt89}, {\it open policies} and {\it closed policies} can be distinguished.
  An open policy (Fig.~\ref{ExempleDesDifferentesPolitiques}.A) expresses all {\it forbidden actions} (called {\it negative authorizations}): a not explicitly denied access is allowed. 
  In a closed policy (Fig.~\ref{ExempleDesDifferentesPolitiques}.B), all {\it authorized actions} have to be fully specified (called {\it positive authorizations}). Finally, some policies are expressed with both positive and negative rules (Fig.~\ref{ExempleDesDifferentesPolitiques}.C). In this case, some actions can be {\it conflicting} or {\it undefined}.

  In the following, readers should distinguish {\it network events}, which are the elementary communication steps of the network, and {\it \B\ events}, which are the description of actions in the \B\ method.

  In the proposed approach, a closed policy defined by a single set ($\SP$) of authorized actions is used. However, each of these abstract actions can be associated to one or more concrete network events and conversely.

\begin{figure}[ht]
    \vskip -16pt
    \centerline{
      \begin{tabular}{c@{~~~~~~}c@{~~~~~~}c}%
        ~~\includegraphics[width=2.6cm]{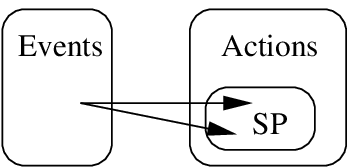}~~
        &~~\includegraphics[width=2.6cm]{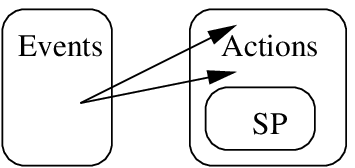}~~
        &~~\includegraphics[width=2.6cm]{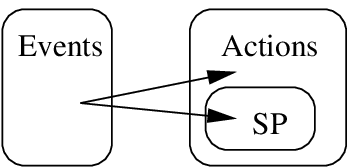}~~\\
        A. Correct with regard to $\SP$
        &B. Violates $\SP$
        &C. In conflict with $\SP$\\
        \small ($\Pass$ status)
        &\small ($\Fail$ status)
        &\small ($\Conflict$ status)\\
      \end{tabular}
    }
    \vskip -7pt
  
    \caption{Events correct with regard to $\SP$, in violation of $\SP$ or in conflict with $\SP$.}
    \label{ExempleDeConformiteConflitViolation}
    \vskip -10pt
  \end{figure}

\begin{definition}[Types of Events]
  An event is \textbf{correct} with regard to a security policy if it corresponds only to authorized actions of the policy  (Fig.~\ref{ExempleDeConformiteConflitViolation}.A).
  If the event is associated only to forbidden actions then this event \textbf{violates} the policy (Fig.~\ref{ExempleDeConformiteConflitViolation}.B).
  If an event is linked to some authorized actions and to some forbidden ones, then this event is in \textbf{conflict} with the policy (Fig.~\ref{ExempleDeConformiteConflitViolation}.C).
\end{definition}

  Several conformity relations~\cite{Jajodia97,Samarati} can be used when conflicting events can occur. 
  The approach proposed in this article is the following:
  
  \begin{definition}[Network Conformity]
    A network conforms to a security policy if 
    each event of this network is correct with respect to the policy.
  \end{definition}

  Finally, if a security policy is relevant only to a part of the network, then all events that are not associated to any action of the policy are {\it unspecified}. 

\vspace*{-3pt}
\section{Policy Security through TCP/IP Levels}
  \subsection{Traceability from Policy to Implementation}
  
  The proposed approach aims to express a security policy at an abstract level (on actions) and to preserve it through the refinement process until its implementation (on network events). However, each refinement level can only access information from the protocol header of the corresponding TCP/IP layer (Fig.~\ref{ModeleTCPIP}) and has to implement the same security policy as the specification.

  The model used to illustrate this approach is a monitor. A monitor has to detect at least each network event which violates $\SP$ or which is in conflict with $\SP$. In the ideal case, no event correct w.r.t. $\SP$ is detected. The monitor has then to guarantee, at each refinement level, the next two properties:
  
  \begin{Propriete}[Monitor Correctness]~\\
    \label{proprieteDeSecurite}
%
      \it Each event that is not detected is correct with respect to the security policy.
  \end{Propriete}

  \begin{Propriete}[Monitor Completeness]~\\
    \label{proprieteDeCompletude}
%
      \it Each event that is detected violates or is in conflict with the security policy.
  \end{Propriete}

  In the model, the network events representation gradually changes at each refinement (from actions to concrete events). To implement these properties, the link between the different representations has then to be modeled, in a systematic way, at each refinement level.
  So, the end user (the administrator of a network) can choose the abstraction level of his policy (by using or not the more abstract levels of the model) and each event is traced through the refinement, as needed for some certification process such as the one of the Common Criteria~\cite{CC}.

  Each refinement level represents how the communication is seen between two elements of the network.
  At level 0, events correspond to the access by a user to a service. They are considered as actions of the security policy. 
  At level 1, events are messages between daemons ({\it Application layer}). 
  At level 2, events are requests between hosts and are attached to particular ports ({\it Transport layer}). 
%
  At level 3, each event is a connection between interfaces and is attached to particular ports ({\it Network layer}). 
  Table~\ref{Tab-CouchesIP} summarizes these different representations.

  \begin{table}[ht]
    \vskip -10pt
    \small
    \begin{center}
      \begin{tabular}{|@{~~~}l@{~~~}|@{~~~}l@{~~~}|@{~~~}l@{~~~}|}
        \hline
        Level of specification     & Network concepts                             & TCP/IP layer\\
        \hline
        \hline
        0 (Policy level, actions)  & Users, services                              & ~ \\
        1                          & Daemons, Terminal servers                    & Application \\
        2                          & Hosts, ports                                 & Transport \\
        3 (implementation)         & Interfaces, ports                            & Network \\
        \hline
      \end{tabular}\\
      \begin{tabular}{l}
        {\it Daemon}: software server providing some services.\\
        {\it Terminal server}: particular daemon providing some logging services.\\
        {\it Host}: machine of the network.\\
        {\it Ports}: channels associated, on each host, to zero or one daemon.\\
        {\it Interface}: network interface (e.g. a network card).\\
      \end{tabular}
    \end{center}

    \vskip -5pt
    \caption{Networks concepts by refinement level}
    \label{Tab-CouchesIP}
    \vskip -20pt
  \end{table}

  These network concepts can be extracted from information contained in configuration files. 
  For example the list of registered Linux users can be found in the \verb!/etc/passwd! file and the list of daemons hosted on each machine can be found in \verb!/etc/init.d/!. This information can then be used to associate each network event to an action of the policy.

  Finally, an observer is introduced in the model to give the internal status ($\Pass$, $\Fail$ or $\Conflict$ - Fig.~\ref{ExempleDeConformiteConflitViolation}) associated to each observed network event. As the event representation changes through the refinement process, the parameters of the observer are described in global variables.

  \subsection{Example}
  
\begin{figure}[ht]
    \vskip -15pt
    \begin{tabular}{cc}
      \parbox{5cm}{
        \centerline{$\!\!$\includegraphics[width=5cm]{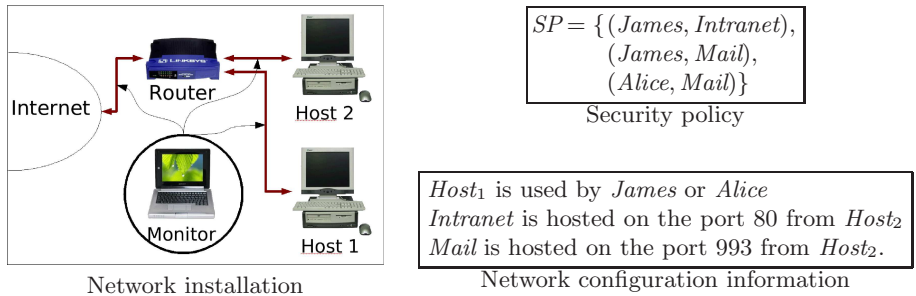}$\!\!$} 
        \centerline{Network installation}
      } & \hspace*{-7pt}\parbox{7.8cm}{
        \machinebox{
          $\!\SP\!=\!\!~\{$\>\>\>\>\>$\!\!\!\!\!(\textit{James}, \textit{Intranet}),$\\
          \>\>\>\>\>$\!\!\!\!\!(\textit{James}, \textit{Mail}),$\\
          \>\>\>\>\>$\!\!\!\!\!(\textit{Alice}, \textit{Mail})\}$
        }
        \centerline{Security policy}
        \leftline{}
        \machinebox{
          $\textit{Host}_1$ is used by $\textit{James}$ or $\textit{Alice}$\\
          $\textit{Intranet}$ is hosted on the port 80 from $\textit{Host}_2$\\
          $\textit{Mail}$ is hosted on the port 993 from $\textit{Host}_2$.
        }
        \centerline{Network configuration information}
      }
    \end{tabular}
  
     \vskip -5pt
    \caption{Example of the monitor installation}
    \label{Exemple}
     \vskip -15pt
  \end{figure}

  To illustrate the notions of conflict and failure, here is a short example (Fig.~\ref{Exemple}) of a monitor that receives a copy of each message from the network and that is parametrized by a security policy and the network configuration. 

  \label{sectionDeLExemple}

  A message coming from $\textit{Host}_1$ and going to $\textit{Host}_2$ at port $80$, is necessarily sent by $\textit{James}$ or $\textit{Alice}$, because they are the only referenced $\textit{Host}_1$ users. Moreover, due to the accessed port of $\textit{Host}_2$, and according to the configuration information, the service can only be the $\textit{Intranet}$. However, the security policy $\SP$ only allows $\textit{James}$ to access to the $\textit{Intranet}$. Thus, the observed message is in $\Conflict$. 

  Now, if the message comes from $\textit{Host}_1$ and goes to port $993$ of $\textit{Host}_2$, then the accessed service is $\textit{Mail}$ and all the users of $\textit{Host}_1$ are authorized to access it. Therefore, this message is correct w.r.t. the security policy ($\Pass$ case).

   Finally, if a message is exchanged with a host ($\textit{Host}_3$) which is not in the described part of the network, then no user or action is associated to it by the network configuration: the event is ignored.

\section{Description of Refinement Levels}

  As previously said, each refinement level represents a different layer of the TCP/IP protocol suite  (Table~\ref{Tab-CouchesIP}). 
  In the first subsection, we define the data domain attached to each refinement level.
  Next, we present a systematic approach to model the links between refinements, based on configuration files. 
  Then, we introduce journals in order to establish the monitor correctness and completeness properties (Properties~\ref{proprieteDeSecurite} and~\ref{proprieteDeCompletude}). 
  Finally, we describe the observer allowing to trace the status of each event through the refinement process.

\subsection{Events Representation}

  Table~\ref{Tab-RepresentationEvent} gives the representation of the events for each refinement level, according to the network concepts presented in Table~\ref{Tab-CouchesIP}. 
  Moreover, the incoming ports are modeled while the outgoing ports are not so (Levels 2 and 3, Table~\ref{Tab-RepresentationEvent}). Outgoing ports are useless as long as the history of connections is not taken into account. Indeed, outgoing ports are dynamically and randomly chosen and cannot be used to identify a daemon or a user, contrary to the incoming ports, that are statically reserved for each service (managed by the IANA\footnote{IANA = Internet Assigned Numbers Authority}).

  \begin{table}
    \vskip -15pt
    \begin{center}
      \begin{tabular}{|@{~~}l@{~~~~}l@{~~}|}
        \hline
        Level of specification    & Network events representation      \\
        \hline
        \hline
        0  {\it (Policy level)}   & $\Event_0 = \USERS \times \SERVICES$\\
        1                         & $\Event_1 = \DAEMONS \times \DAEMONS$\\
        2                         & $\Event_2 = \HOSTS \times (\HOSTS \times \PORTS)$\\
        3  {\it (Implementation)} & $\Event_3 = \INTERFACES \times (\INTERFACES \times \PORTS)$\\
        \hline
      \end{tabular}
    \end{center}
    \vskip -5pt
    \caption{Network events representation for each refinement level}
    \label{Tab-RepresentationEvent}
    \vskip -25pt
  \end{table}

  The policy is enforced only on the known part of the network. The constant  $\ConfigurationDomain_i$ represents the known network subset at each level $i$. Each set is divided into a known and an unknown part, as follows:

  \vspace*{1pt}
  \begin{ConstantsProperties}{Known Network Definition}
    \small
    \hspace*{-4pt}\begin{tabular}{l}
      $\Users \!\subset\! \USERS \!\!\logand\!\! \Services \!\subset\! \SERVICES \!\!\logand\!\! \Daemons \!\subset\! \DAEMONS \!\!\logand\!\! \Hosts \!\subset\! \HOSTS$\\
      $\!\!\logand\! \TerminalServers \subseteq \Daemons \!\logand\! \Ports \subset \PORTS \!\logand\! \Interfaces \subset \INTERFACES$\vspace*{5pt}
    \end{tabular}\\

    \hspace*{-4pt}\begin{tabular}{l}
      $\!\!\logand\! \ConfigurationDomain_0 ~=~ \Users \times \Services$\\
      $\!\!\logand\! \ConfigurationDomain_1 ~=~ \TerminalServers \times \Daemons$\\
      $\!\!\logand\! \ConfigurationDomain_2 ~=~ \Hosts \times (\Hosts \times \Ports)$\\
      $\!\!\logand\! \ConfigurationDomain_3 ~=~ \Interfaces \times (\Interfaces \times \Ports)$
    \end{tabular}
  \end{ConstantsProperties}
  \vspace*{1pt}

  These abstract sets correspond to concrete data extracted from configuration files. 
  Finally, the security policy is described by the constant set $\SP$ of all authorized actions of the known network.

  \vspace*{4pt}
  \begin{ConstantsProperties}{Security Policy Definition}%
      \small $\SP \subseteq \ConfigurationDomain_0$\\
  \end{ConstantsProperties}
  \vspace*{-4pt}

\subsection{Representation Relation}
\label{SectionRefines}
  
  The event representation changes through the refinement process. The relation $\Refines_i$ defines the representation link between the $i^{th}$ and $(i-1)^{th}$ refinement levels. 
  For example, $\Refines_1$ associates each terminal server to a set of users and each daemon to a set of services. Note that $\Refines_i$ is a relation and not a function because a concrete event is not always associated to a single abstract event (as seen in the example from Section~\ref{sectionDeLExemple}). It is neither a function from $\ConfigurationDomain_{i-1}$ to $\ConfigurationDomain_i$ because an action can be associated to several concrete events. 
  %
  These relations can be composed to define $\Refines_{i\rightsquigarrow0}$ between the i$^{th}$ level and the policy level. 
  
  \begin{ConstantsProperties}{Representation Relation axioms}
    \label{InvariantRefines}
      \small $\Refines_i \in \ConfigurationDomain_i \brel \ConfigurationDomain_{i-1}$ &~~~~ (with $i \!\in\! 1..3$)\\ 
      \small $\!\!\!\logand\!\! \Refines_{i\rightsquigarrow0} \!\in\! \ConfigurationDomain_{i} \brel \ConfigurationDomain_0$                                &~~~~ (with $i \!\in\! 1..3$)\\
      \small $\!\!\!\logand\!\! \Refines_{i\rightsquigarrow0} = (\Refines_i;\Refines_{i-1};\dots;\Refines_1)$ &~~~~ (with $i \!\in\! 1..3$)
  \end{ConstantsProperties}

  In order to simplify some further invariants, we define $\Refines_0$ as the identity ($\rid(\Event_0)$).
  Finally, each element mentioned in the configuration has to be associated with at least one action and one network event:

  \begin{ConstantsProperties}{The Described Sub-Network is Known as a Whole}
    \label{InvariantLaContrainte}
        \small $\rdom(\Refines_i) = \ConfigurationDomain_i ~~\logand~~ \rran(\Refines_i) = \ConfigurationDomain_{i-1}$\\
  \end{ConstantsProperties}

  \vspace*{-15pt}
\subsection{Journalizing Observed Events}
\label{MiseEnPlaceJournaux}
  \vspace*{-2pt}

  Three journals are maintained: the one of observed events ($\Monitored_i$) and two other ones of warned events ($\FAIL_i$ and $\CONFLICT_i$ respectively for the events which violate and are in conflict with the policy). All these journals are defined as non-ordered sets, because the considered policy does not take into account the history. Moreover, all warned events are observed and all events observed in a concrete level are also observed in the abstract level. At the policy level, no conflict can occur, then $\CONFLICT_0$ is empty.

  \begin{Invariant}{General Journals Definition} 
    \label{InvariantJournaux}
        \small $\Monitored_0 \subseteq \ConfigurationDomain_0$ \\
        \small $\!\!\logand \Monitored_i \subseteq \Refines_i^{-1}[\Monitored_{i-1}]$ ~~&~~~~~ (with $i \in 1..3$)\\
        \small $\!\!\logand \CONFLICT_0 = \emptyset$ \\
        \small $\!\!\logand \FAIL_i \cup \CONFLICT_i \subseteq \Monitored_i$ &~~~~~ (with $i \in 0..3$)\\
        \small $\!\!\logand \FAIL_i \cap \CONFLICT_i = \emptyset$ &~~~~~ (with $i \in 0..3$)\\
  \end{Invariant}

  \noindent Properties~\ref{proprieteDeSecurite} and~\ref{proprieteDeCompletude} can be expressed on these journals, by the next two invariants:

  \noindent (1) All observed events associated with $\Pass$ status are correct w.r.t. $\SP$:

  
  \begin{Invariant}{Monitor Correctness - Property~\ref{proprieteDeSecurite}} 
    \label{InvariantProp1}
        \small $\Refines_{i\rightsquigarrow0}[\Monitored_i-\FAIL_i-\CONFLICT_i] \subseteq \SP$&~~~~~~~ (with $i \in 0..3$)\\
  \end{Invariant}

  \noindent (2) No event associated with $\Conflict$ or $\Fail$ status is correct w.r.t. $\SP$: 
  
  \begin{Invariant}{Monitor Completeness - Property~\ref{proprieteDeCompletude}}
    \label{InvariantProp2}
    
    \small $\Refines_{i\rightsquigarrow0}[\FAIL_i] \cap \SP =\emptyset$&~~~~~~~ (with $i \in 0..3$)\\
    \small $\!\!\logand \forall e_i \cdot (e_i \in \CONFLICT_i \bimpl \Refines_{i\rightsquigarrow0}[\{e_i\}] \not\subseteq \SP)$&~~~~~~~ (with $i \in 0..3$)\\
  \end{Invariant}

  \vspace*{-13pt}
  \subsection{Observer Introduction}
\label{SectionObservateur}  
  \vspace*{-2pt}
  
  The \B\ event $\getstatus$ is an observer of the network events. It returns the status of an event chosen in a nondeterministic way. 
  Because of the change of event representation, the observer is modeled with two new global variables: $\EVar$ and $\ERes$. $\EVar$ is the chosen observed event ($\EVar_i\in \ConfigurationDomain_i$) and $\ERes$ is its status ($\ERes_i \in \{\Pass, \Fail, \Conflict\}$).
  Fig.~\ref{GetStatusi} gives the general implementation of the observer $\getstatus$.

  \begin{figure}[ht]
    \machinebox{
      \\$\getstatus \bdef$ \bany $e_i$ \bwhere $e_i \in \Monitored_i$ \bthen\\
      \>\bif $e_i \in \FAIL_i$        \bthen\>~~~~~~~~~~~~~~~~~~~~~~~~~~~~~~~~~~~~~ $\EVar_i:=e_i ~||~\ERes_i := \Fail$\\
      \>\belsif $e_i \in \CONFLICT_i$ \bthen\>~~~~~~~~~~~~~~~~~~~~~~~~~~~~~~~~~~~~~ $\EVar_i:=e_i ~||~\ERes_i := \Conflict$\\
      \>\belse                              \>~~~~~~~~~~~~~~~~~~~~~~~~~~~~~~~~~~~~~ $\EVar_i:=e_i ~||~\ERes_i := \Pass$\\
      \>\bend\\
      \bend
    }
    \vskip -7pt
    \caption{General definition of the $\getstatus$ event}
    \label{GetStatusi} 
    \vspace*{-10pt}
  \end{figure}

  The variables $\EVar$ and $\ERes$ are defined at each refinement level with the following invariant:
%
%
%
%
%
  \begin{Invariant}{Observed Variables}
    \label{InvariantObservedVariables0}
    \label{InvariantObservedVariables123}
      \small $\Monitored_i\not=\emptyset \bimpl $\\
      \small $~~~~~~~~((\ERes_i=\Fail)                     \logequiv (\EVar_i\in\FAIL_i))$\\
      \small $~~~~~~~~\!\!\logand ((\ERes_i=\Conflict)                 \logequiv (\EVar_i\in\CONFLICT_i))$\\
      \small $~~~~~~~~\!\!\logand ((\ERes_i=\Pass)                     \logequiv (\EVar_i\in \Monitored_i-\FAIL_i-\CONFLICT_i))$\\
  \end{Invariant}
  
  The observed event is traced through the refinement process:
  
  \begin{Invariant}{Relation through Refinement}
    \small $(\EVar_i, \EVar_{i-1}) \in \Refines_i$
    \label{InvariantObservedVariablesThroughRefinement}
  \end{Invariant}

  Finally, correctness and completeness of the monitor (Properties~\ref{proprieteDeSecurite} and~\ref{proprieteDeCompletude}) are implemented on the journals with the invariants defined in Section~5.3. However, these properties can also be checked on the observed variables.
  So, if the following assertions hold, then Properties~\ref{proprieteDeSecurite} and~\ref{proprieteDeCompletude} are verified:

  \begin{Assertion}{Monitor Correctness on Observed Variables}
  \label{AssertionProp1}
      \small $\Monitored_i\not=\emptyset \logand \ERes_i=\Pass     \bimpl \ERes_{i-1}=\Pass $&~~~~~~~ (with $i \in 0..3$)
  \end{Assertion}

  \begin{Assertion}{Monitor Completeness on Observed Variables}
  \label{AssertionProp2}
    \small $\Monitored_i\not=\emptyset \bimpl $& (with $i \in 0..3$)\\
    \small $~~~~~~(\ERes_i=\Fail     \bimpl \ERes_{i-1}=\Fail)$\\
    \small $~~~~~~\!\!\!\logand\!\! (\ERes_i\!=\!\Conflict \!\bimpl\! \Refines_{i\rightsquigarrow0}[\{\EVar_i\}]\!\not \subseteq\! \SP)$\\
    \small $~~~~~~\!\!\!\logand\!\! (\ERes_i\!=\!\Conflict \!\bimpl\! \Refines_{i\rightsquigarrow0}[\{\EVar_i\}]\!\cap\! \SP\not\!=\!\emptyset )$\\
  \end{Assertion}

  Therefore we only discuss the verification of those assertions that are sufficient to show Properties~\ref{proprieteDeSecurite} and~\ref{proprieteDeCompletude}.

\section{Model Description}

  In the previous section, we have presented, in a systematic way, all data required for the model development.
  In this section, we describe successively each level by introducing: the configuration data, the \B\ events and the construction of the $\Refines_i$ relation. All invariants and properties described in the previous section are included at each description level. 
  
  In this description, we also focus on the verification of the correctness and the completeness properties of the monitor by checking Assertions~\ref{AssertionProp1} and~\ref{AssertionProp2}.

\subsection{Level 0: User-Service View}

   The $\SP$ constant set is given by the user while constants $\Users$ and $\Services$ are retrieved from configuration files. 
%
%
  Journals are represented as abstract variables and are empty in the initial state 
%
%
%
%
%
%
%
%
%
%
  ($\Monitored_0 := \emptyset$ and $\FAIL_0 := \emptyset$). Consequently, the observed variables are initially undefined ($\EVar_0 :\in\Event_0 \wedge \ERes_0:\in\{\Pass,\Fail,\Conflict\}$). 
If an event $e_0$ occurs on the network then: 
  \begin{itemize}
   \item if $e_0$ is not in the observed sub-network then $\eventfilter$ (Figure~\ref{EventFilter0}.B) is launched and $e_0$ is ignored, 
   \item else $\checkevent$ (Figure~\ref{CheckEvent0}.A) is launched and $e_0$ is stored in $\Monitored_0$. Moreover, if $e_0$ violates the policy then it is journalized in $\FAIL_0$.
  \end{itemize}

  \begin{figure}[ht]
    \vskip -25pt
    \begin{center}
      \begin{tabular}{ccc}
        \machinesbox{
          $\checkevent \bdef$\bany $e_0$ \bwhere $e_0 \in \ConfigurationDomain_0$ \bthen \\
          \>$\Monitored_0 := \Monitored_0 \cup \{e_0\} ~||$\\
          \>\bif $e_0 \not\in \SP$ \bthen\ $\FAIL_0:=\FAIL_0 \cup \{e_0\}$ \bend\\
          \bend
        }& &\machinesbox{
          $\eventfilter \bdef$\bany $e_0$\\
          \>\bwhere $e_0 \in \Event_0$\\ 
          \>\>$\!\!\logand e_0 \not\in \ConfigurationDomain_0$\\
          \>\bthen\ $\bskip$ \bend
        }\\
        A. $\checkevent$ & & B. $\eventfilter$\\
      \end{tabular}
    \end{center}
    \vskip -20pt
    \caption{Code of \B\ events $\checkevent$ and $\eventfilter$ at the policy level.}
    \label{CheckEvent0} 
    \label{EventFilter0} 
    \label{GetStatus0} 
    \vskip -15pt
  \end{figure}

  \noindent Finally, this level establishes Properties~\ref{proprieteDeSecurite} and~\ref{proprieteDeCompletude} by verifying Assertions~\ref{AssertionProp1} and~\ref{AssertionProp2} with the observed variables only, since there is no conflict at this level.

\subsection{Level 1: Servers View}

  According to Table~\ref{Tab-CouchesIP}, the daemons sets ($\Daemons$ and $\TerminalServers$) are now described. The relation between this level and the more abstract one, i.e. the policy level, is extracted from configuration files. Each daemon is configured with its registered users and its provided services. 
  We model this information with two relations ($\provide$ and $\usedby$) describing which user can be connected to a particular terminal server and which daemon provides a particular service.
  For example, the registered users list of a telnet server can be found in \verb!/etc/passwd!. 

  \begin{ConstantsProperties}{$\Refines_1$ Definition}
    \small
    \begin{tabular}{c}
     $\usedby \in \TerminalServers \brel \Users ~~\logand~~ \provide \in \Daemons \brel \Services$  \\
     $\!\!\logand\Refines_1 = (\usedby ~||~ \provide)$  \\
    \end{tabular}
  \end{ConstantsProperties}

  \begin{figure}[ht]
    \vskip -25pt
    \small
    \begin{center}
      \begin{tabular}{ccc}
        \machinesbox{
          $\checkevent \bdef$ \bany $e_1$ \bwhere $e_1 \in \ConfigurationDomain_1$ \bthen \\
          \>$\Monitored_1 := \Monitored_1 \cup \{e_1\} ~||$\\
          \>\blet $E_0$ \bbe $E_0 = (\usedby ~||~ \provide)[e_1]$ \bin\\
          \>\>\bif\ $E_0 \cap \SP \not= \emptyset$ \bthen\\
          \>\>\>$\FAIL_1:=\FAIL_1 \cup \{e_1\}$\\
          \>\>\belsif\ $E_0 \not\subseteq \SP$ \bthen\\
          \>\>\>$\CONFLICT_1:=\CONFLICT_1 \cup \{e_1\}$\\
          \>\>\bend\\
          \>\bend\\
          \bend
        }& &\machinesbox{
          $\eventfilter \bdef$\bany $e_1$\\
          \>\bwhere $e_1 \in \Event_1$\\ 
          \>\>$\!\!\logand e_1 \not\in \ConfigurationDomain_1$\\
          \>\bthen\ $\bskip$ \bend
        }\\
        A. $\checkevent$ & & B. $\eventfilter$
      \end{tabular}
    \end{center}
    \vskip -20pt
    \normalsize
    \caption{Code of the \B\ events $\checkevent$ and $\eventfilter$ at level 1.}
    \label{CheckEvent1} 
    \label{EventFilter1} 
    \vskip -15pt
  \end{figure}

  The journals and the observed variables are defined according to the invariants given in Section~5. 
  At this level of refinement, the relation $\Refines_1$ is used dynamically by the \B\ event $\checkevent$ (Figure~\ref{CheckEvent1}.A) to compute the status of the observed network event, while the \B\ event $\eventfilter$ (Figure~\ref{EventFilter1}.B) ignores all messages exchanged with the unknown part of the network.

  The observer refinement (Fig.~\ref{GetStatusi}) produces 18 proof obligations, which, associated to the three ones generated for Assertions~\ref{AssertionProp1} and~\ref{AssertionProp2}, establish Properties~\ref{proprieteDeSecurite} and~\ref{proprieteDeCompletude} on the model.

\subsection{Level 2: Hosts View}

  As described in Table~\ref{Tab-CouchesIP}, this level introduces the notions of $\Hosts$ and $\Ports$. 
  The relation between these concepts and the daemons is extracted from hosts configuration information. They are summarized by two functions: $\hosting$, which associates the hosts to the daemons, and $\runon$, which precises the ports used by a particular daemon on a host. 
  Configuration data is such that:

  \begin{ConstantsProperties}{$\Refines_2$ Definition}
    \small
      \begin{tabular}{c}
        $\hosting \in \Hosts \brel \Daemons ~~\logand~~ \runon \in \Hosts \times \Ports \bpar \Daemons$\\
        $\!\!\logand \Refines_2 = (\hosting \brestcodom \TerminalServers) ~||~ \runon$\\
      \end{tabular}
  \end{ConstantsProperties}

  Just as in previous levels, the journals and the observed variables are defined according to the invariants given in Section~5. The relation $\Refines_2$ is used by $\checkevent$ (Figure~\ref{CheckEvent2}.A) to compute the status of the observed network event, while $\eventfilter$ (Figure~\ref{EventFilter2}.B) ignores all messages exchanged with the unknown part of the network.
%
%
%
%
 
  The main contribution of this level is the implementation of the $\checkevent$ event (Figure~\ref{CheckEvent2}.A), which progressively refines the method to compute the $E_0$ set of all actions associated to the observed network events $e_2$.

  \begin{figure}[ht]
    \vskip -17pt
    \small
    \machinebox{
      \\$\checkevent \bdef$ \bany $e_2$ \bwhere $e_2 \in \ConfigurationDomain_2$ \bthen \\
      \>$\Monitored_2 := \Monitored_2 \cup \{e_2\} ~||$\\
      \>\blet $E_1$ \bbe $E_1 = ((\hosting \brestcodom \TerminalServers) ~||~ \runon)[\{e_2\}] $ \bin\\
      \>\>\blet $E_0$ \bbe $E_0 = (\usedby ~||~ \provide)[E_1]$ \bin\\
      \>\>\>\bif\ $E_0 \cap \SP \not= \emptyset$ \bthen\\
      \>\>\>\>$\FAIL_2:=\FAIL_2 \cup \{e_2\}$\\
      \>\>\>\belsif\ $E_0 \not\subseteq \SP$ \bthen\\
      \>\>\>\>$\CONFLICT_2:=\CONFLICT_2 \cup \{e_2\}$\\
      \>\>\>\bend\\
      \>\>\bend\\
      \>\bend\\
      \bend
    }
    \centerline{A. $\checkevent$}
    \machinebox{
      $\eventfilter \bdef$\bany $e_2$ \bwhere $e_1 \in \Event_2-\ConfigurationDomain_2$ \bthen\ $\bskip$ \bend
    }
    \centerline{B. $\eventfilter$}
    \vskip -10pt
    \normalsize
    \caption{Code of the \B\ events $\checkevent$ and $\eventfilter$ at level 2.}
    \label{CheckEvent2} 
    \label{EventFilter2} 
    \vskip -17pt
  \end{figure}

  In the same way as in the previous level, Properties~\ref{proprieteDeSecurite} and~\ref{proprieteDeCompletude} are established by proving the three proof obligations generated for Assertions~\ref{AssertionProp1} and~\ref{AssertionProp2} and the 32 proof obligations generated for the observer event $\getstatus$.

\subsection{Level 3: Implementation}
    \vskip -4pt

  At this level, hosts are valuated into their IP address (32~bit natural which identifies hosts for the Network layer) and ports remain unchanged. For example, the host anchieta.imag.fr can be valuated into its IP address $129.88.39.37$ by using its 32~bit natural value\footnote{$2170038053 = ((129*256+88)*256+39)*256+37$}: $2170038053$. 
  
  The $\Refines_3$ relation is thus the identity and all invariants are inherited and do not need to be proven again. All other constants (security policy and configuration information) have also to be valuated. 
  Network parameters can be retrieved in configuration files while the security policy has to be given by the administrator.
  Table~\ref{SourceDonnees} gives some concrete examples for Fedora-Core (a Linux distribution) of files containing usable data. 

   \begin{table}[ht]
    \vskip -18pt
     \small
     \begin{center}
       \begin{tabular}{|@{~~}l@{~~}|@{~~}l@{~~}|@{~~}l@{~~}|}
         \hline
         Refinement level & Constant & Data file\\
         \hline
         \hline
         0 (Policy level) & $\Users$ and $\Services$ & \verb!/etc/passwd! and \verb!/etc/init.d/!\\
         1 & $\provide$ and $\usedby$ & Configuration files of each server\\
         2 & $\runon$ and $\hosting$& \verb!/etc/services! and \verb!/etc/init.d/!\\
         3 (Implementation) & $\Interfaces$ & \verb!/etc/hosts!\\
         \hline
       \end{tabular}
     \end{center}
     
     \normalsize
     \vskip -7pt
     \caption{Example of configuration files for Fedora-Core system.}
     \label{SourceDonnees}
    \vskip -28pt
   \end{table}

  However, the event-\B\ language cannot be directly implemented. The model is translated into classical \B. This transformation is done by some ad-hoc methods based on the results of the MATISSE\footnote{Methodologies and Technologies for Industrial Strength Systems Engineering (MATISSE): IST Programme RTD Research Project (2000-2003).} project \cite{Matisse,EvtB2B}. 
  Since the guards of $\eventfilter$ and $\checkevent$ are disjoint, the events are replaced by the single operation $\checkevent\_and\_\eventfilter$ (Fig.~\ref{CheckEvent3}).
  Moreover, the network event $e_3$ chosen in the guard \bany $e_3$ \bwhere $e_3 \in \Event_3$ is replaced by three input parameters $\mathit{IP}_1$, $\mathit{IP}_2$ and $Po$ representing respectively the IP address of the two hosts and the incoming port. 

  \begin{figure}[ht]
    \small
    \vskip -17pt
    \machinebox{
      \\$\checkevent\_and\_\eventfilter (\mathit{IP}_1,\mathit{IP}_2,Po) \bdef$ \bbegin\\
      \>\>{\it /* Typing precondition: }$(\mathit{IP}_1,(\mathit{IP}_2,Po)) \in \Event_3$ {\it */}\\
      \>\bvar $tmp$ \bin\\
      \>\>$tmp \bres \isInConfDom (\mathit{IP}_1,\mathit{IP}_2,Po)~;$\\
      \>\>\bif $tmp=${\sc true} \bthen  \>\>\>\>~~~~~~~~~~~~~~~~~~~~~~~~~~~~~~~~~~~~~~~~~~~~~~~~~~~~~~~~~{\it /* Case of $\checkevent$ */}\\
      \>\>\>$Src:=\mathit{IP}_1~;~Dest:=\mathit{IP}_2~;~Port:=Po~;$\\
      \>\>\>$tmp \bres \isoutofSP (\mathit{IP}_1,\mathit{IP}_2,Po)~;$\\
      \>\>\>\bif\ $tmp = ${\sc true} \bthen\ $\Status:=\Fail~;\WriteFail(\mathit{IP}_1,\mathit{IP}_2,Po)$\\
      \>\>\>\belse\\
      \>\>\>\>$tmp \bres \isinSP (\mathit{IP}_1,\mathit{IP}_2,Po)~;$\\
      \>\>\>\>\bif\ $tmp = ${\sc false} \bthen\ $\Status:=\Conflict~;\WriteConflict(\mathit{IP}_1,\mathit{IP}_2,Po)$\\
      \>\>\>\>\belse $\Status:=\Pass$ \bend\\
      \>\>\>\bend\\
      \>\>\bend                     \>\>\>\>~~~~~~~~~~~~~~~~~~~~~~~~~~~~~~~~~~~~~~~~~~~~~{\it /* Else case of $\eventfilter$: $\bskip$ */} \\
      \>\bend\\
      \bend
    }
    \vskip -10pt
    \caption{Implementation of the B event $\checkevent$.}
    \label{CheckEvent3} 
    \vskip -5pt
  \end{figure}

  Figure~\ref{CheckEvent3} is the implementation of $\checkevent\_and\_\eventfilter$ operation. It uses three local operations ({\small $\isInConfDom$}, {\small $\isoutofSP$} and {\small $\isinSP$}) to compute the correctness of each observed event. These operations are implemented as refinements of the corresponding parts of $\checkevent$ at level 2. For example, the local operation $\isoutofSP$ is defined in Figure~\ref{isoutofSPCODE}.
  
  \definecolor{gris}{rgb}{.3,.3,.3}  
  \begin{figure}[ht]
    \vskip -12pt
    \machinebox{
    \\\small$rr \bres \isoutofSP(\mathit{IP}_1,\mathit{IP}_2,Po) \bdef$\\
    \small\>\bpre $(\mathit{IP}_1,(\mathit{IP}_2,Po)) \in \ConfigurationDomain_3$ \bthen\\
    \small\>\>$rr \!=\! \bbool((\usedby ~||~ \provide)[$~~~~~~~~~~~~~~~~~~~~~~~~~~~~~~~~\textcolor{gris}{{\it /*}$\Refines_1 [${\it */}}\\
    \small\>\>~~~~~~~~~~~~~~~$((\hosting \!\brestcodom\! \TerminalServers) ~||~ \runon)[$~~~~\textcolor{gris}{{\it /*}$~\Refines_2 [${\it */}}\\
    \small\>\>~~~~~~~~~~~~~~~~~~~$\{(\mathit{IP}_1,(\mathit{IP}_2,Po))\}$~~~~~~~~~~~~~~~~~~~~~~~~~~~~~~\textcolor{gris}{{\it /*}$~~\{(\mathit{IP}_1,(\mathit{IP}_2,Po))\}${\it */}}\\
    \small\>\>~~~~~~~~~~~~~~~~$]$~~~~~~~~~~~~~~~~~~~~~~~~~~~~~~~~~~~~~~~~~~~~~~~~~~~~~~~~\textcolor{gris}{{\it /*}$~~]${\it */}}\\
    \small\>\>~~~~~~~~~~~~~$]\cap \SP = \emptyset)$~~~~~~~~~~~~~~~~~~~~~~~~~~~~~~~~~~~~~~~~~~~~~\textcolor{gris}{{\it /*}$~]\cap \SP = \emptyset${\it */}}\\
    \small\>\bend
    }
    \vskip -10pt
    \caption{Abstract definition of the $\isoutofSP$ local operation.}
    \label{isoutofSPCODE}
    \vskip -12pt
  \end{figure}

  The $\Monitored_i$ set, modeled to store the monitored events, is not implemented, while $\FAIL$ and $\CONFLICT$ sets are stored in files. That is managed by an external component providing the operations {\small $\WriteFail$} and {\small $\WriteConflict$}.
  However, Properties~\ref{proprieteDeSecurite} and~\ref{proprieteDeCompletude} still hold, since the observed variable $\ERes_2$ and $\EVar_2$  remain unchanged. 

\begin{figure}
  \vskip -8pt
  \centering
  \includegraphics[width=8.5cm]{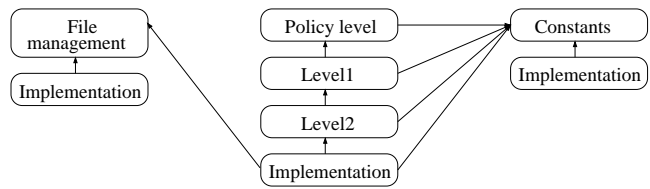}
  \vskip -8pt
  \caption{General model organisation}
  \label{ExporationConstantes}
  \vskip -40pt
\end{figure}

  Finally, constants (configuration data and security policy) are exported in an external component as shown in Fig.~\ref{ExporationConstantes}. Thus, the model is generic and can be completely proved independently of the configuration data. The user just needs to provide some network information, or to retrieve it from configuration files, and 
  to fulfill the conditions on configuration stated in Sections~5.3 and 5.2.
  A similar approach has been used in M\'et\'eor~\cite{Meteor}, the Parisian subway without driver, to develop some generic and reusable components.

  If the model is valuated for a simple example of network with two hosts, two daemons, one service and one user, then 31 proof obligations are generated and only 26 of them are discharged by the automatic prover. 
  The five remaining proof obligations have been interactively discharged, but are really obvious and only need two commands : replace (\verb!eh!) and predicate prover (\verb!pp(rp.0)!).

%

\section{Conclusion}

  This work has been done within the framework of the POTESTAT\footnote{
    Security policies: test directed analysis of open networks systems.
  } project \cite{potestat}, which aims at proposing a methodology for network security testing from high-level security policies.
  The main problem is to establish the conformity relation in order to automatically generate test cases and oracles from an abstract specification, as it has been implemented in the TGV tool~\cite{tgv} from IRISA and Verimag french laboratories.

  Our contribution is a method to automatically enforce an abstract security policy on a network. 
  In order to achieve that, we build a formal relation ($\Refines_{i\rightsquigarrow0}$) between abstract and concrete levels. 
  The dynamic part of the program ($\checkevent\_and\_\eventfilter$) computes all actions associated to each observed event. 
  Finally, we guarantee, by using an observer ($\getstatus$), that all violations and conflicts are detected at each refinement level (Property~\ref{proprieteDeSecurite}) and that no warning is issued for an event which is correct with respect to the policy (Property~\ref{proprieteDeCompletude}).

  The work of D.~Senn, D.~Basin and of G.~Caronni~\cite{Basin} and of G.~Vigna~\cite{Vigna} are also dealing with the modeling of network conformity of a security policy. The model of~\cite{Vigna} supports all the TCP/IP layers but does not provide a formal definition of the policy and needs human interaction to produce the test cases, while the model introduced in~\cite{Basin} only considers the first two layers and uses a low-level policy.

  In this paper, we described the development of a network monitor, and the same approach can be used for  generation, verification or test of a network configuration. The relationship is built in the same way and only the dynamic part of the model has to be modified.

  In particular, many works have been developed relative to the generation of firewall configurations. For example, {\sf Firmato}~\cite{Bartal99} is a tool generating the firewall configuration from a security policy and a network topology, and the {\sf POWER} tool~\cite{Power}, of Hewlett-Packard, can rewrite a security policy into devices configuration. However, {\sf Firmato} needs some topology information and uses a low-level policy, and {\sf POWER} requires some human interactions during the process. 
  The work presented here 
  does not need human interaction (if all proof obligations are automatically discharged) or topology information. Due to the existence of the conflict status, it seems more adapted to the monitoring approach.

  Finally, in our model, the conflict case can be removed if $\Refines_{i\rightsquigarrow0}$ is a function from $\ConfigurationDomain_i$ to $\ConfigurationDomain_0$, as done in~\cite{Cuppens04} with a security policy expressed in the OrBAC framework~\cite{OrBac}. 
%
  In order to achieve that, we have to recognize the user and the service associated to each event. 
  It can be realistic to associate only one service to each port, but it is too strict to impose that each host can be used by only one user. 
  An investigation should be done to properly compare their approach with ours.

\vspace*{-10pt}
\subsubsection*{Acknowledgments.}
The authors would like to thank  
D.~Bert, V.~Darmaillacq, V.~Untz, F.~Dadeau and Y.~Grunenberger for their advises and their reviews.
\vspace*{-10pt}

\bibliographystyle{plain}
\bibliography{Bibliographie-B,Bibliographie-SecuReseau}

\end{document}